# CALORIC MICRO-COOLING: Numerical modelling and parametric investigation

Jan Kalizan, Jaka Tušek*

Faculty of Mechanical Engineering, University of Ljubljana, Aškerčeva 6, 1000 Ljubljana, Slovenia

*e-mail: jaka.tusek@fs.uni-lj.si

## Abstract

Cooling systems based on the caloric effects of ferroic materials show high potential for various cooling and heat-pumping applications due to their potentially high efficiencies and the lack of any environmentally hazardous refrigerants. One of such applications that has recently gained the attention of the scientific community is micro-cooling, which can be applied for hot-spot cooling and thermal management in electronic components. In this study a comprehensive numerical analysis of a caloric micro-cooling system using elastocaloric and electrocaloric materials was performed in order to investigate the limits and potential of this technology. We demonstrated that a caloric micro-cooling system is able to cool down the electronic component below room temperature or at least stabilize it at lower temperatures compared to the case when only the heat sink is applied in an efficient way (with the COP values exceeding 10). The specific cooling capacity of the caloric micro-cooling device strongly depends on the heat sink accompanied with the caloric material and its heat transfer capabilities. The caloric device can cool the electronic component below room temperature at heat-flux densities of up to 0.35 W/cm$^2$ and up to 1 W/cm$^2$ if it is used together with air-cooled heat sinks and water-cooled heat sinks, respectively. Caloric cooling systems could therefore play an important role as an efficient micro-cooling technology for certain applications, in particular where under-cooling below room temperature (low-temperature electronics) is required.

**Highlights**:

- A new 1D transient caloric micro-cooling numerical model was developed
- Parametric investigation was performed with different caloric materials
- Elastocaloric elements can generate heat-flux densities up to 1 $W/cm^2$
- Electrocaloric elements allows highly efficient operation with COP well above 10

## 1. Introduction

The thermal management of electronic components is one of the most critical issues in the electronics industry [1]. As the power and packing density of electronic components increases [2], the amount of waste heat generated in a small space rises significantly. In some applications it is expected to exceed 300 W/cm$^2$ in the near future [3]. The failure rate of electronic equipment increases exponentially with temperature and generally needs to be maintained below 350 K [3]. Also, the high thermal stresses as a result of temperature differences in electronic components are major causes of failure. As a result, thermal control has become increasingly important in the design and operation of electronic equipment. Furthermore, in certain applications, such as complementary metal-oxide semiconductors (CMOSs), lasers, and infrared detectors, it is desirable or even necessary that they operate in a below-room-temperature regime where active cooling (i.e., refrigeration) is required. However, depending on the power density and the application of an electronic component, different cooling mechanisms and systems are available. They can range from air-cooled heat sinks with extended surfaces and forced convection, to liquid-cooling systems with single-phase forced convection or evaporation, to heat pipes and thermoelectric (i.e., Peltier) coolers, etc. A comprehensive review of the cooling of electronic components can be found in e.g., [3], [4], [5]. With forced air-cooling the effective heat-transfer coefficients can be up to 300 W/m$^2$K, which results in a heat flux of up to approximately 2 W/cm$^2$ (at a temperature difference of 80 K). Single-phase water cooling can have effective heat-transfer coefficients up to 3000 W/m$^2$K, which results in a heat flux of up to around 25 W/cm$^2$ (at a temperature difference of 80 K), while with evaporation the heat-transfer coefficients can be as high as 100,000 W/m$^2$K with a heat flux above 500 W/cm$^2$ (at a temperature difference of 80 K) [3]. The heat fluxes can be increased even up to 1000 W/cm$^2$ if micro-channels are employed and/or nanofluids are used instead of conventional fluids [6], [7]. Heat pipes have attracted a lot of attention from the electronics industry due to their high performance (high effective thermal conductivity and low thermal resistance), an absence of moving parts and no external power consumption [8], [9], [10]. Micro-channel heat pipes can dissipate the heat fluxes of electronic components in the range 100 to 1000 W/cm$^2$ [9]. Thermoelectric coolers can enhance the cooling power density (when used in a combination with air or liquid heat sinks) [11], [12], [13], they also have no moving parts, they can be miniaturized and have the possibility to be of cooling below ambient temperature, but their efficiency is generally low (COP below 1 when the temperature span exceed 20 K [14]).

However, recently, new mini/micro-cooling systems based on the caloric effects in ferroic materials were purposed. The caloric effect refers to the magnetocaloric, electrocaloric, elastocaloric and barocaloric effects and has recently attracted attention as an alternative cooling or heat-pumping mechanism due to its potentially high efficiency and the lack of environmentally hazardous refrigerants [15], [16]. When an external field (magnetic, electric or mechanical) is applied to the appropriate ferroic material, a solid-state transformation occurs in the material and latent heat is released from the material (or under adiabatic conditions the material heats up). When an external field is removed a reverse process occurs and the latent heat is absorbed by the material (or under adiabatic conditions the material cools down), which can be applied in cooling or heat-pumping applications. Materials exhibit multiple-coupled caloric effects with more than one external field are called multicaloric materials [17]. In recent years several proof-of-concept devices and some theoretical analyses of miniaturized electrocaloric and elastocaloric systems have been described. In general, caloric devices can be classified as two types: having a regenerative operational principle with convective heat transfer (e.g., active magnetic regenerator normally used in magnetic refrigeration [18], [19]) or a single-stage operational principle (or a cascade system with several single-stage elements connected in series) with a contact-to-contact heat transfer that allows for miniaturization [20], [21], [22]. Due to external constraints (e.g., bulky magnetic field sources) the magnetocaloric effect is not considered as a potential micro-cooling mechanism, while the barocaloric effect (although showing high potential [23]), was not yet demonstrated in a proof-of-concept device. Therefore, to date only the electrocaloric and elastocaloric effects have demonstrated the potential for micro-cooling applications. Most of the miniature devices are designed as single-stage devices with a single caloric plate in between a heat sink and heat source, as shown schematically in Figure 1. The basic operation of such a single-stage cooling system is as follows. In the first step a caloric plate is subjected to an external field (electric field or mechanical stress), so it heats up. In the second step the caloric plate is in a thermal contact with the heat sink, where heat is transferred to the surroundings. In the third step the caloric plate is removed from the external field, so it cools down; and in the fourth step the caloric plate is in a thermal contact with the heat source (electronic component), where the heat is absorbed by the caloric plate. A thermal contact can be provided by the movement of the caloric plate to/from the heat sink/heat source or by the implementation of a thermal diode or a thermal switch, which can actively control the direction and magnitude of the heat flux. A thermal diode/switch in a caloric cooling system needs to provide the highest possible thermal resistance between the surfaces where there should be no heat transfer (polarized/loaded caloric plate and heat source or depolarized/unloaded caloric plate and heat sink) and the highest heat transfer between the surfaces to which the heat flux is transferred (polarized/loaded caloric plate and heat sink or depolarized/unloaded caloric plate and heat source). For more details about thermal diodes and thermal switchers see [24], [25].

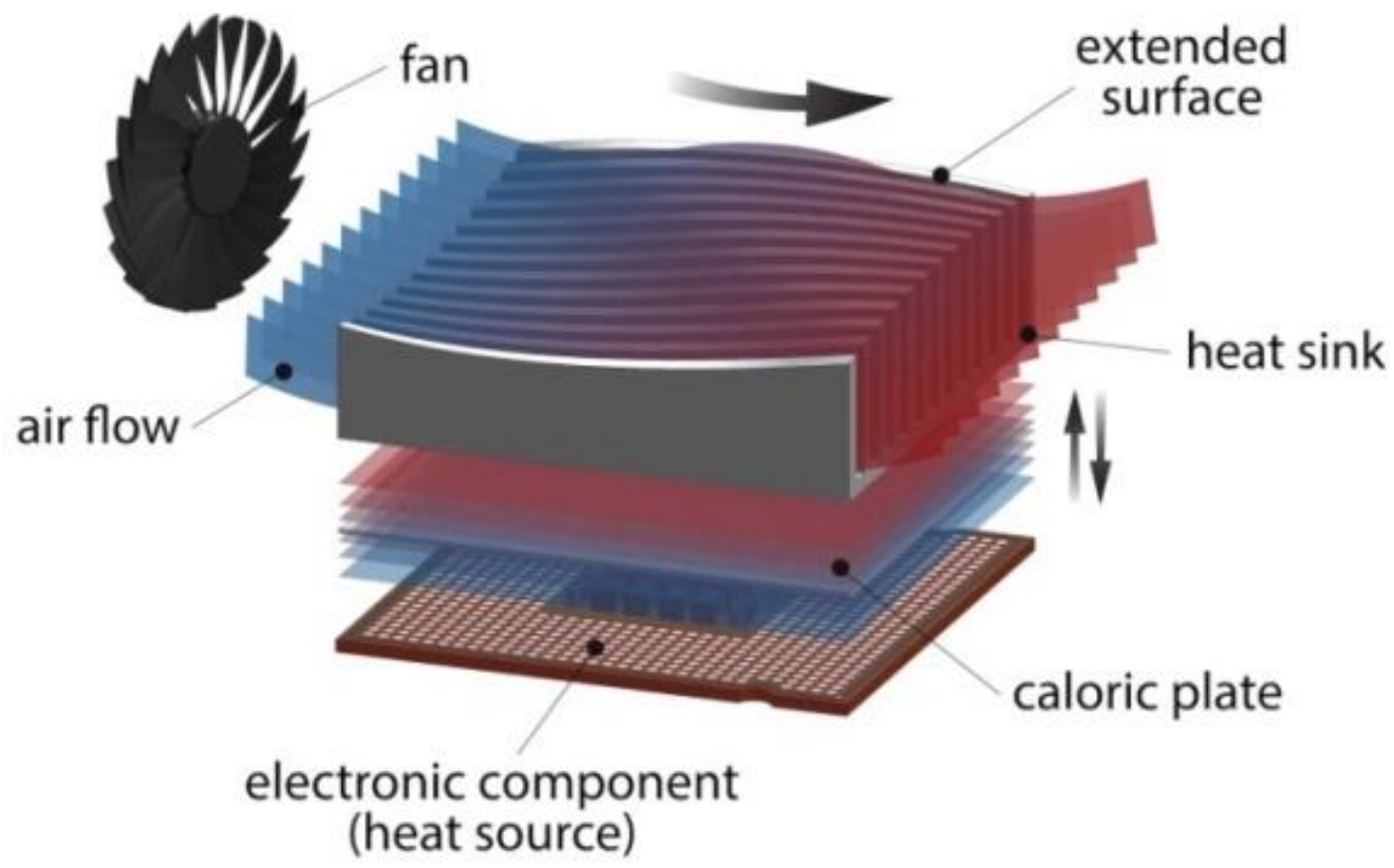

Figure 1: Schematic presentation of a caloric micro-cooling device.

The first single-stage elastocaloric device was proposed and analysed in 1969 by Lawless [26] for cryogenic applications. In 1988, Basiulis and Barry [27] proposed a cascade system with heat pipes working as thermal diodes in between each electrocaloric element. Similarly, Karmanenko et al. [28] simulated the performance of cascade electrocaloric elements working as a solid-state cooling line based on alternative adiabatic and isothermal switching of the electrocaloric elements that made it possible to generate a directed heat flux and a temperature gradient along the structure. Epstein and Malloy [28] theoretically investigated a single-stage electrocaloric device with a heat switcher that provides thermal contact between the electrocaloric material and the heat sink/source at appropriate times. They concluded that if the conductivity ratio of a heat switcher exceeds 100, the Carnot efficiency of such a type of device can be above 66%. Ju [29] theoretically evaluated a single-stage cooling device based on the electrocaloric effect for electronics cooling and showed that if the electrocaloric material with an adiabatic temperature change of 10 K operates with a frequency above 10 Hz it can produce a specific cooling power of above 10 W/cm$^2$. In 2011, Casasanta [30] proposed a device where the movement of an electrocaloric material and therefore the thermal contact with the heat sink/source is driven by electrostatic forces. The first single-stage electrocaloric device was built and tested in 2012 by Jia and Ju [31]. It was based on doped $BaTiO_3$ ferroelectric layers in a multilayer capacitor, which was mechanically actuated up and down to ensure alternating thermal contacts with a heat source/sink. The largest temperature span between the heat sink and the heat source measured for this device was 1 K (at an operating frequency of 0.3 Hz). In 2013, Gu et al. [32] presented a chip-scale electrocaloric cooling device based on a sold-state regenerator, which is analogous to an active electrocaloric regenerator [33], where instead of a heat-transfer fluid a sold matrix is used to transfer the heat between the electrocaloric material and the heat sink/source, which can increase the compactness of the whole device. As the electrocaloric material, a multi-layered P(VDF-TrFE-CFE) polymer was used. The largest temperature span measured on this device was 6.6 K (at an operating frequency of 1 Hz and an applied electric field of 100 MV/m). Gu et al. [34] have also theoretically investigated this type of electrocaloric device and showed that it can provide 9 W/cm$^3$ of volumetric specific cooling power for a 20 K temperature span and a Carnot efficiency up to 50%. In 2015 the first single-stage electrocaloric device using thermal diodes was built and tested by Wang et al. [35]. As an electrocaloric material, a multi-layered $BaTiO_3$ ceramic material was used, while silicon thermal switchers (see [36] for details) were used as the heat-transfer mechanism. They measured a maximum cooling power of 36 mW for a temperature span of 0.3 K (with an electric field change of 28 MV/m). Smullin et al. [37] have further simulated the performance of this type of device and they showed that it can generate a specific cooling power of 30 mW/cm$^2$ for a zero temperature span. In 2016, Hirasawa et al. [38] proposed and analysed numerically a single-stage electrocaloric device with a thermal switcher based on the fluid motion between the electrocaloric film and the heat sink/source. Bradeško et al. [39] proposed a new cantilever-based cascade cooling system using a relaxor ferroelectric material that has a double function in such a system. It works as a refrigerant (based on the electrocaloric effect), as well as a thermal switch, since it provides the thermal contact for the heat transfer between the cantilever and the heat sink/source at the appropriate times (due to the electromechanical response). They showed theoretically that 15 of such cantilevers in a cascade system can achieve 12.6 K of temperature span. In 2017, Ma et al. [40] designed, built and tested a single-stage electrocaloric device using a two-layered P(VDF-TrFE-CFE) flexible electrocaloric polymer film and an electrostatic actuation mechanism. The device produced a specific cooling power of 2.8 W/g and a COP of 13. A similar electrostatic actuation mechanism applied as a heat-transfer mechanism in a caloric film was studied in [41]. Recently, Sun et al. [42] proposed and theoretically evaluated a single-stage electrocaloric device based on flexural deformation due to the converse piezoelectric effect, which enables the device to bend in order to ensure thermal contact with the heat source/sink.

Miniaturized elastocaloric systems are currently less explored in comparison with electrocalorics. The first single-stage elastocaloric experimental set-up was designed and built in 2015, where the mechanical travel of an elastocaloric element (Ni-Ti plate) was used to ensure an alternating thermal contact with the heat sink/source [43]. The largest temperature span between the heat sink and the heat source measured in this device was around 8 K. More recently, Ossmer et al. [44] and Bruederlin et al. [45] developed and tested a series of miniaturized single-stage elastocaloric devices. Their latest device is designed as an antagonistic pair connecting two elastocaloric plates, where one is loaded and the other one simultaneously unloaded in order to recover some energy released during the unloading and to increase the overall device's efficiency. The operation is based on the actuation of elastocaloric elements between a flat heat source and a convex heat sink. When the elastocaloric element is in contact with the heat source it is released (and therefor cooled due to the elastocaloric effect), while when it is contact with the heat sink it is loaded over a convex geometry and therefore heats up due to the elastocaloric effect. They have tested different Ni-Ti-based materials (Ni-Ti-Fe and Ni-Ti-Cu-Co) under different operating conditions. The largest temperature span measured between the heat sink and the heat source was about 14 K; the largest specific cooling power was 18 W/g; and the largest COP value was up to 6 [45]. The have also simulated the performance of this type of device and obtained a specific cooling power of 0.125 W/cm$^2$ [45]. Furthermore, they have demonstrated the possibility of cascading this type of device to enhance its performance [46].

Nevertheless, even though miniaturized cooling systems based on the electrocaloric and elastocaloric effects show real potential for the efficient thermal management of electronic components and hot-spot cooling, no systematic analyses of a caloric micro-cooling device's operation have been performed to date. The main aim of this paper is a comprehensive theoretical (numerical) analysis and investigation/optimization of a realistic caloric micro-cooling device and its operating conditions with different caloric materials in order to demonstrate the potential and limitations of this novel technology.

## 2. Methods

### *2.1.Mathematical model*

A new, one-dimensional and time-dependent numerical model for the simulation and optimization of a single-stage micro-cooling system has been developed. It is based on an electrocaloric or elastocaloric element working as an active body that transfers heat from an electronic component in which heat flux is generated, through a heat sink, to the ambient (see Figure 1). The model is built around a governing equation based on the law of energy conservation in a control volume [47] of the system under consideration. As the system (Figure 1) is highly symmetrical and the side areas of the system are negligibly smaller than the frontal areas, the model can be reduced to a one-dimensional form. Figure 2 shows the one-dimensional governing equations of the system, where three different parts of the system (i.e., heat sink, caloric plate and heat source) with different governing equations and different boundary conditions can be distinguished. The heat stored in the control volume of all three elements of the system (left-hand side of Eqs. (2), (5) and (8)) depends on the density ($\rho$) of the material, its specific heat ($c$) and the transient temperature change over time ($\partial T/\partial t$). The heat conduction depends on the temperature profile over a control volume in the relevant direction ($\partial^2 T/\partial x^2$) and the thermal conductivity of the material ($k$). For a caloric plate, there is an additional term on the right-hand side of Eq. (5) that describes the caloric effect. As explained later in the text, this term is only applied in the loading/polarization and unloading/depolarization steps of the cooling cycle. The term ($\partial s/\partial X$) describes the partial derivative of the isothermal entropy change of the caloric effect over the applied strain/polarization as a function of strain/polarization and temperature, while the term $\partial X/\partial t$ describes the strain/polarization rate of the caloric material ($X$ represents the internal variable of the caloric effect, i.e., deformation/polarization). This is the most accurate and the most widely applied approach of introducing the caloric effect into the governing differential equation of a caloric materials and was applied in various numerical models of different caloric devices [20], [33], [48]. The governing equation of the heat source (Eq. (8)) also includes a specific heat-flux generation term ($q_{gen}$) that is equal to the specific power of the electronic component. At the lower border of the system ($x=x5$) an adiabatic condition is assumed, while at the upper border ($x=0$) the heat is transferred to the ambient by heat convection through an extended surface with a forced convection to enhance the effective heat-transfer coefficient ($h_{eff}$). The internal boundary conditions between the caloric plate and the heat sink/source (Eqs. (3), (4), (6), (7)) depend on the step of the cooling cycle. When the caloric plate is in contact with the heat sink/source, a contact heat transfer with contact thermal resistance ($R$) occurs, while on the other side of the caloric plate a complete thermal isolation with an adiabatic boundary condition is assumed at that time.

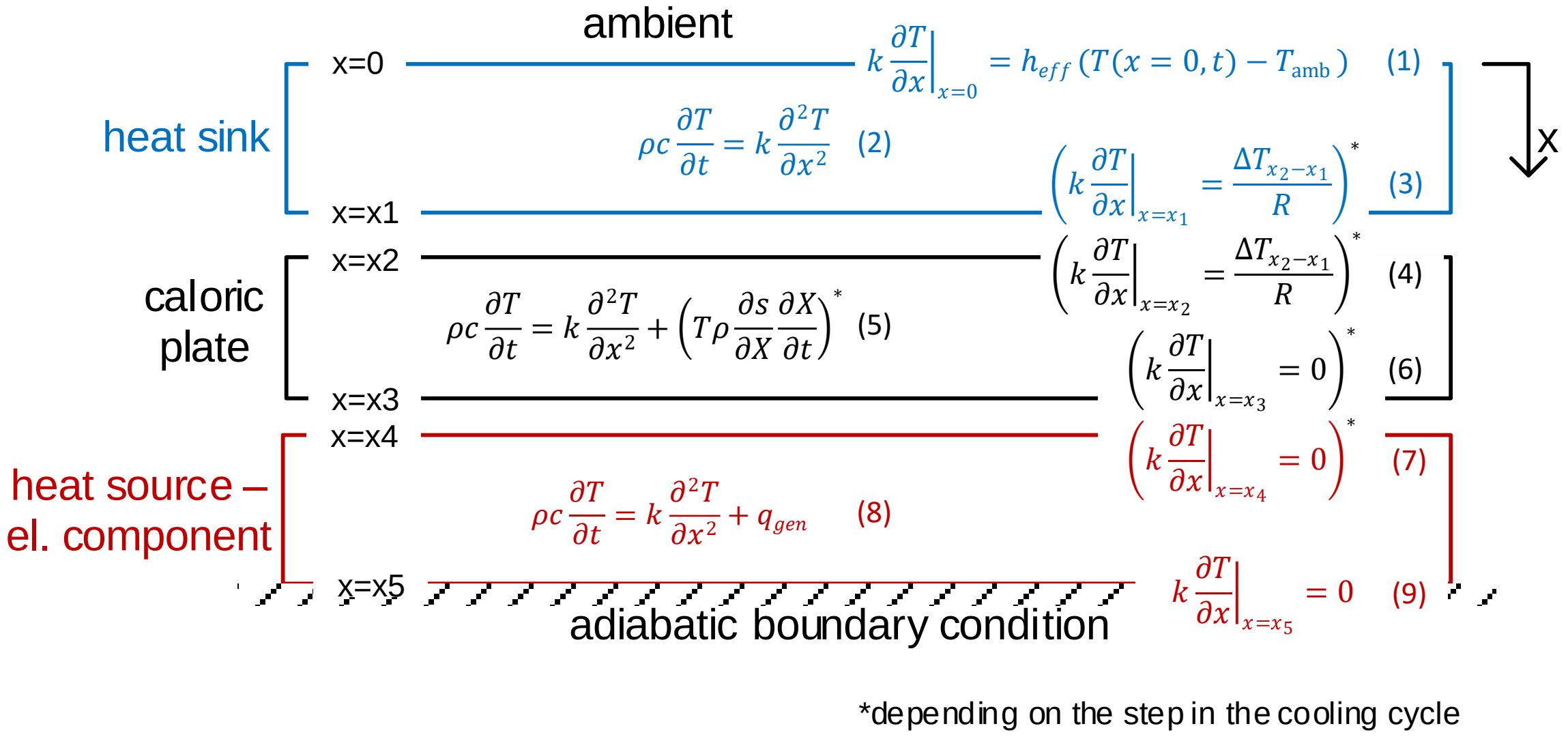

Figure 2: Set of governing equations and related boundary conditions applied in the numerical model of a caloric micro-cooling system.

The evaluated cooling process is based on the Brayton thermodynamic cycle, which is the most common caloric cooling cycle, although other thermodynamic cycles can also be applied [49], [50]. The Brayton thermodynamic cycle is based on four operational steps (see Figure 3):

- Loading/polarization: The caloric plate is not in thermal contact with either the heat sink or the heat source (adiabatic boundary conditions are applied on both sides of the caloric plate). It heats up due to the caloric effect (the term describing the caloric effect is included in the governing equation of the caloric plate – Eq. (5)).
- Heat transfer to the heat sink: The caloric plate is in thermal contact with the heat sink and isolated from the heat source (contact heat transfer is applied between the caloric plate and the heat sink, while the adiabatic boundary condition is applied to the heat source). In this step the caloric plate remains loaded/polarized (the term describing the caloric effect is not included in the governing equation of the caloric plate).

- Unloading/depolarization: The caloric plate is not in thermal contact with either the heat sink or the heat source (adiabatic boundary conditions are applied on both sides of the caloric plate). It cools down due to the caloric effect (the term describing the caloric effect is included in the governing equation of the caloric plate).
- Heat transfer from the heat source: The caloric plate is in thermal contact with the heat source and isolated from the heat sink (contact heat transfer is applied between the caloric plate and the heat source, while the adiabatic boundary condition is applied to the heat sink). In this step the caloric plate remains unloaded/depolarized (the term describing the caloric effect is not included in the governing equation of the caloric plate).

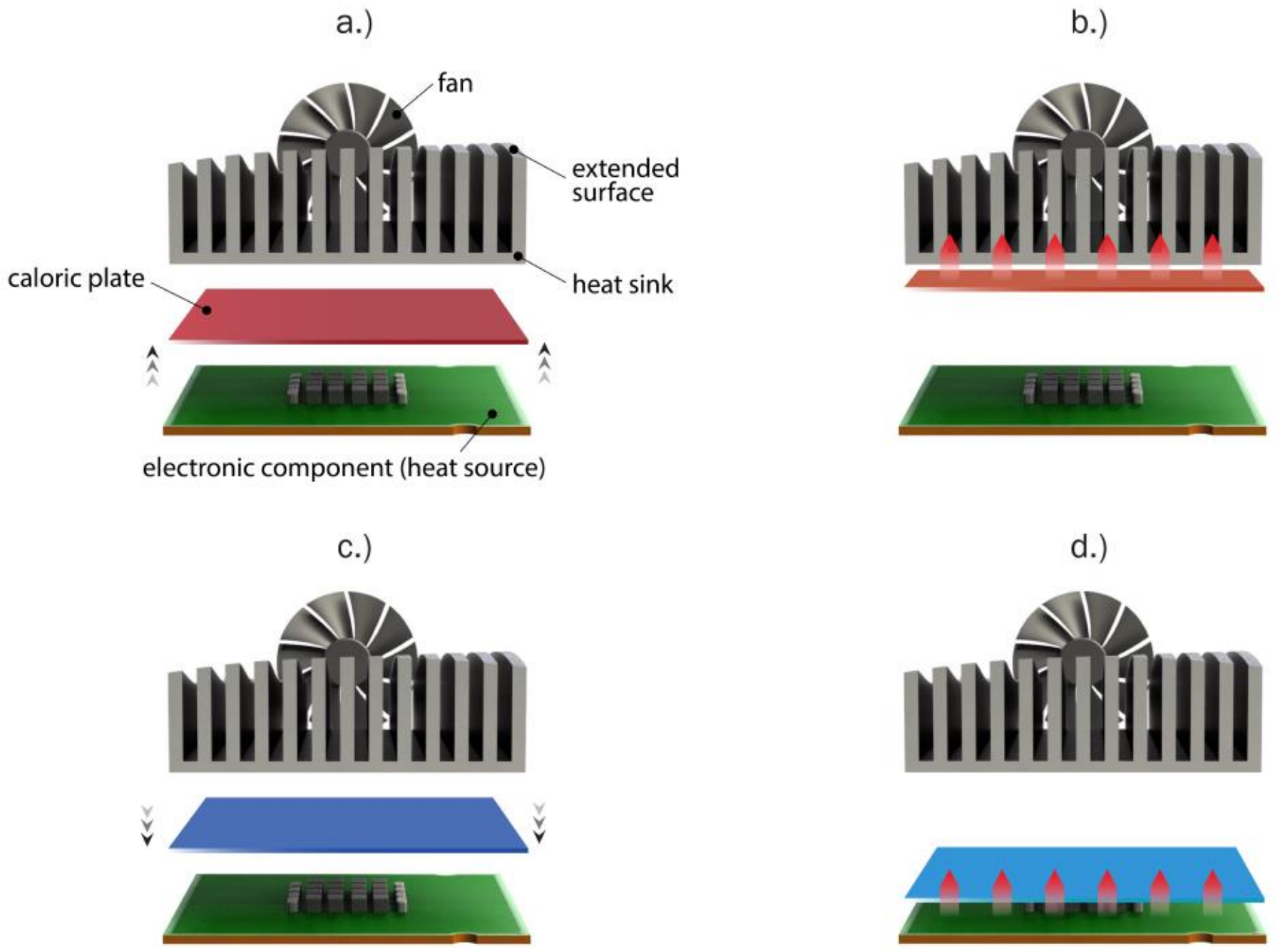

Figure 3: Schematic presentation of four operational steps of a caloric micro-cooling device (Brayton thermodynamic cycle): a.) Loading/polarization; b.) Heat transfer to the heat sink; c.) Unloading/depolarization; d.) Heat transfer from the heat source.

## 2.2. The caloric properties

In this study three different caloric materials were evaluated: one elastocaloric material, i.e., the Ni-Ti alloy; and two electrocaloric materials, i.e. the [Pb(Mg1/3Nb2/3)O3] 0.9[PbTiO3]0.1 (PMN-10PT) relaxor ceramic and the P(VDT-TrFE-CFE) polymer, which currently show one of the highest potential for practical applications and are the most widely applied elastocaloric [22] and electrocaloric materials [32], [33], [40], respectively. All the required caloric properties of the materials evaluated in the numerical model (specific heat ($c$) and partial derivative of the isothermal entropy change over the applied strain/polarization ($\partial s/\partial X$) as a function of the strain/polarization and temperature) were, in general, obtained based on experimentally measured adiabatic temperature changes ($\Delta T_{ad}$). The Ni-Ti alloy was evaluated at two applied strains. The high strain (6%) corresponds to the end of the transformation plateau with a complete martensitic transformation [51], [48], while the small strain (2%) corresponds to the fatigue-resistant operation of a thin Ni-Ti sheet in tension [52]. The electrocaloric materials were evaluated at intermediate electric fields in order to prevent a potential electrical breakdown, Joule heating [53] and/or fatigue [54], [55]. Therefore, the electrocaloric properties of the PMN-10PT relaxor ceramic and the P(VDT-TrFE-CFE) polymer evaluated in this study were obtained at 53 kV/cm [33] and at 750 kV/cm [56], respectively.

In order to ensure the thermodynamic consistency of the caloric data set, a previously developed phenomenological model (that can predict the behaviour of ferroic materials) in a combination with the Maxwell relations [48], [57] was applied to fit the experimentally obtained adiabatic temperature changes (see Figure 4). As seen in Figure 4, for the three cases, (Ni-Ti at 2%, PMN-10PT and P(VDT-TrFE-CFE)), the modelled adiabatic temperature changes fit well with the experimental one. On the other hand, the modelled values of the adiabatic temperature changes of the Ni-Ti alloy strained at 6% is significantly over-predicted.

The reasons for this over-prediction are described in [51] and can be mainly attributed to the non-complete transformation during the adiabatic temperature-change experiments, while a complete transformation was assumed in the applied phenomenological model. Therefore, in order to demonstrate the full potential of the caloric material during the complete transformation, the modelled values of the caloric properties were used for further modelling of the device (and not directly fitted ones).

The following Maxwell equations were used for the calculation of the partial derivative of the isothermal entropy change over the applied strain/polarization:

$$\frac{\partial T_{ad}}{\partial Y} = \frac{T}{c}\frac{\partial Y}{\partial T} \leftrightarrow \Delta T_{ad} = \int_{Y_1}^{Y_2} \frac{T}{c}\frac{\partial Y}{\partial T} dY \tag{10}$$

$$\frac{\partial s}{\partial X} = \frac{\partial Y}{\partial T} \leftrightarrow \Delta s = \int_{Y_1}^{Y_2} \frac{\partial Y}{\partial T} dY \tag{11}$$

where $X$ is an internal variable (polarization or strain) that is changed due to the application of an external field $Y$ (electric field or mechanical stress). The specific heat ($c$) of the caloric material around its transformation (Curie) temperature strongly depends on the temperature and the applied field, and was calculated with the following equation:

$$c = T\frac{\partial S_{tot}}{\partial T} \tag{12}$$

where $S_{tot}$ is the total entropy as a function of the temperature and the applied external field. The first step towards a calculation of the total entropy is to calculate its value at zero applied field using the following equation:

$$S_{tot,Y=0} = \left(\int_{T_1}^{T_2} \frac{c_0(T_1)}{T} dT\right) + \Delta s(Y = Y_{max}) \tag{13}$$

where $c_0$ represents the baseline specific heat (at the reference temperature $T_1$), which can be taken as a constant value (see Table 1). This value can either be obtained by calorimetry measurements or taken from the literature. The $\Delta s(Y = Y_{max})$ is the isothermal entropy change at the final (maximum applied) electric field / mechanical stress (and as a function of temperature). The temperature $T_1$ is a reference temperature at which the total entropy is assumed to be zero. The reference temperature should be well below the transformation (Curie) temperatures at which the total entropy is independent of the external field and the caloric effect equals zero. The total entropy for different applied external fields can be further calculated by adding isothermal entropy changes ($\Delta s$) at different temperatures and applied external fields:

$$S_{tot} = S_{tot,Y=0} + \Delta s \tag{14}$$

All the evaluated caloric materials in this study exhibit hysteresis behaviour, which results in the temperature irreversibility between a positive adiabatic temperature change during the external field application and the negative adiabatic temperature changes during the field removal (Figure 4). It is therefore important to include these irreversibilities in the model as they can significantly decrease the efficiency of the device. If the material shows hysteresis behaviour, the calculation of the caloric properties shown above (Eqs. (10) - (14)) is only valid for the application of the external field, while for the field removal the hysteresis irreversibility needs to be taken into account. The irreversibilities of all the caloric materials where calculated indirectly, by calculating the hysteresis area of a stress-strain or polarization-electric field loop (Eq. (15)) and are presented in Table 1.

$$\Delta s_{hyst} = \frac{1}{\rho \cdot T} \oint X dY \tag{15}$$

The entropy irreversibility calculated with Eq. (15) was further subtracted from the isothermal entropy change calculated with Eq. (11) to obtain the isothermal entropy change for the field removal process. With this approach a different set of caloric properties ($\partial s/\partial X$) was obtained for the external field application and removal (a similar approach was previously demonstrated in [33], [48], [58]) in order to include the hysteresis losses in the caloric device's model.

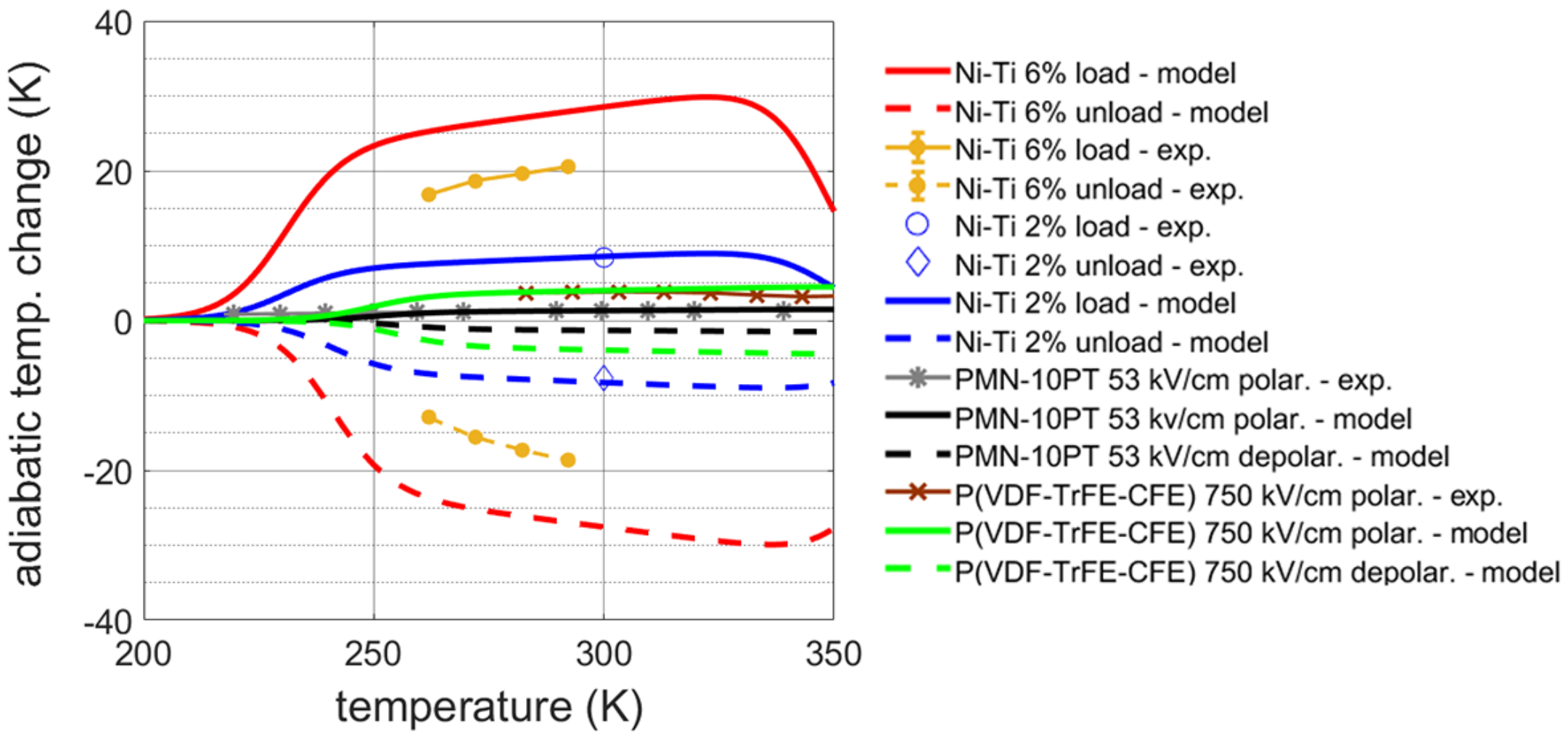

Figure 4: Experimental and modelled values of the adiabatic temperature changes as a function of the temperature for the caloric materials under consideration in this study.

### *2.3.Numerical model*

Since the mathematical model (Eqs. (1) - (9)) does not have an adequate exact analytical solution numerical methods were applied. The governing equations and the boundary conditions (Eqs. (1) - (9)) shown in Figure 2 were discretized using the finite-difference method where the implicit Euler method was used for the discretization of the time-dependent derivatives. The obtained set of algebraic equations were written in the matrix equation $\left([A]\cdot[\vec{T}]=[\vec{B}]\right)$, where $A$ is a tridiagonal matrix of coefficients obtained from discretized algebraic equations, $T$ is the vector of the calculated temperatures and $B$ is the vector of known values (the temperatures of a previous time step and boundary conditions). The discretization and detailed matrix composition are shown in the Appendix A. Supplementary data. The matrix equation is further implemented in the Matlab software and solved for each time step during each of four processes in the cycle until the steady-state conditions are achieved. In the steady-state conditions the software calculates the temperature of the electronic component ($T_{el}$), the temperature span between the heat sink and the source ($\Delta T$), and the COP of the cycle, which is defined as:

$$COP = \frac{q_{gen}}{w_{in}} \tag{16}$$

where $q_{gen}$ is the specific heat flux density generated by the electronic component and $w_{in}$ is the input work required to run a thermodynamic cycle induced by an external field (including the hysteresis losses) – see Eq. (17). Since the electronic component acts as the heat source, its specific heat-flux density essentially presents the specific cooling capacity generated by the cooling device. For the purposes of the calculation of the input work it was assumed that the energy released during the unloading/depolarization process can be fully recovered. The energy consumption related to the heat sink (fan) is not considered in the COP calculation.

The flow chart of the program is presented in Figure 5. The discretization parameters ($m$ and $n$) and steady-state condition criterion ($e$) where high/low enough to ensure a low discretization error (below 0.1%). Before its use the numerical model was also verified for the thermodynamic consistency by comparing the input work required to run the cooling cycle (after the periodic steady state conditions were reached) calculated by two independent approaches:

$$w_{in} = \underbrace{q_{hhex} - q_{gen}}_{w_{in1}} = \underbrace{\rho \cdot d \cdot f \oint T ds}_{w_{in2}} \tag{17}$$

where $\rho$, $d$, $f$, $T$ and $s$ are density of the caloric material, thickness of the caloric plate, operating frequency, temperature and entropy, respectively. In Eq. (17) $w_{in1}$ is calculated as a difference of the specific heat flux transferred through the heat sink to the ambient ($q_{hhex}$) and the specific heat flux generated by the electronic component ($q_{gen}$), while $w_{in2}$ is calculated based on the enclosed area of the performed thermodynamic cycle in the T-s diagram. Note that $w_{in2}$ was used as $w_{in}$ in Eq. (16). However, as show in detail in Figure S14 in the Appendix A.

Supplementary data both input works calculated with Eq. (17) agrees well and their deviation does not exceed 5% at a wide range of operating conditions. The deviations can be attributed to numerical errors and errors due to possible incompatibility of the caloric data ($\partial s/\partial X$ and $c$). The model can therefore be considered at thermodynamically consistent and correct (with estimated accuracy of ±5 %). Similar verification method is often applied in numerical models of active magnetic regenerators in magnetocaloric refrigeration technology [20].

Table 1: Properties of materials used in the simulations.

| | Ni-Ti<br>[51], [48], [52] | PMN-10PT<br>[33], [59] | P(VDT-TrFE-CFE)<br>[56] | Copper |
|---|---|---|---|---|
| $\rho$ (kg/m$^3$) | 6500 | 8130 | 1800 | 8930 |
| $c$ (J/kgK) | 480* | 350* | 1500* | 383 |
| $k$ (W/mK) | 10 | 1.3 | 0.2 | 395 |
| $\Delta s_{hyst}$ (J/kgK) | 1.207 (for 6%)<br>0.362 (for 2%) | 0.0013 | 0.25 | n.a. |

*baseline specific heat – real values are calculated with Eq. (12)

Table 2: Initial, boundary and operating conditions used in the simulations.

| | |
|---|---|
| $T_{surr}$ (K) | 290 |
| $h_{eff}$ (W/m$^2$K) | 200–1000 |
| $R$ (m$^2$K/W) | 2x10$^{-4}$–10$^{-5}$ |
| $\nu$ (Hz) | 0.1–2.5 |
| $P_{load}$ (s) | 0.1 |
| $P_{transf}$ (s) | 0.1–5 |
| $d_c$ (mm) | 0.1–5 |
| $d_{hs}$ (mm) | 5 |

Table 1 shows the values of the material's properties for all three caloric materials and copper, which was applied as a heat sink/source material. As noted, the specific heat values of the caloric materials presented in Table 1 are baseline specific heat values, while its exact values, as a function of the temperature and the applied external field, were calculated using Eq. (12). Initial, boundary and operating conditions are shown in Table 2. The ambient and the initial temperature ($T_{surr}$) were assumed to be 290 K. The effective heat-transfer coefficient ($h_{eff}$) is a product of the heat-transfer coefficient and the fin effectiveness (the ratio of the heat transferred from the finned surface and the heat transferred from the same surface without fins) [47], while $R$ is the contact thermal resistance. Both the effective heat-transfer coefficient and the contact thermal resistance were varied in the simulations to evaluate its impact on the cooling performance of the device. The operating frequency was varied between 0.1 and 2.5 Hz. The applied Brayton thermodynamic cycle includes two different processes, i.e., the application/removal of the external electric field or stress period ($P_{load}$) and the period of the heat transfer ($P_{transf}$). The $P_{load}$ should be as short as possible (to increase the operating frequency) and was assumed to be 0.1 s, while the $P_{trasnf}$ was varied in order to find the optimum value. The thickness of the caloric plate ($d_c$) was varied between 0.1 and 5 mm, while the thickness of heat sink/source ($d_{hs}$) was kept constant at 5 mm.

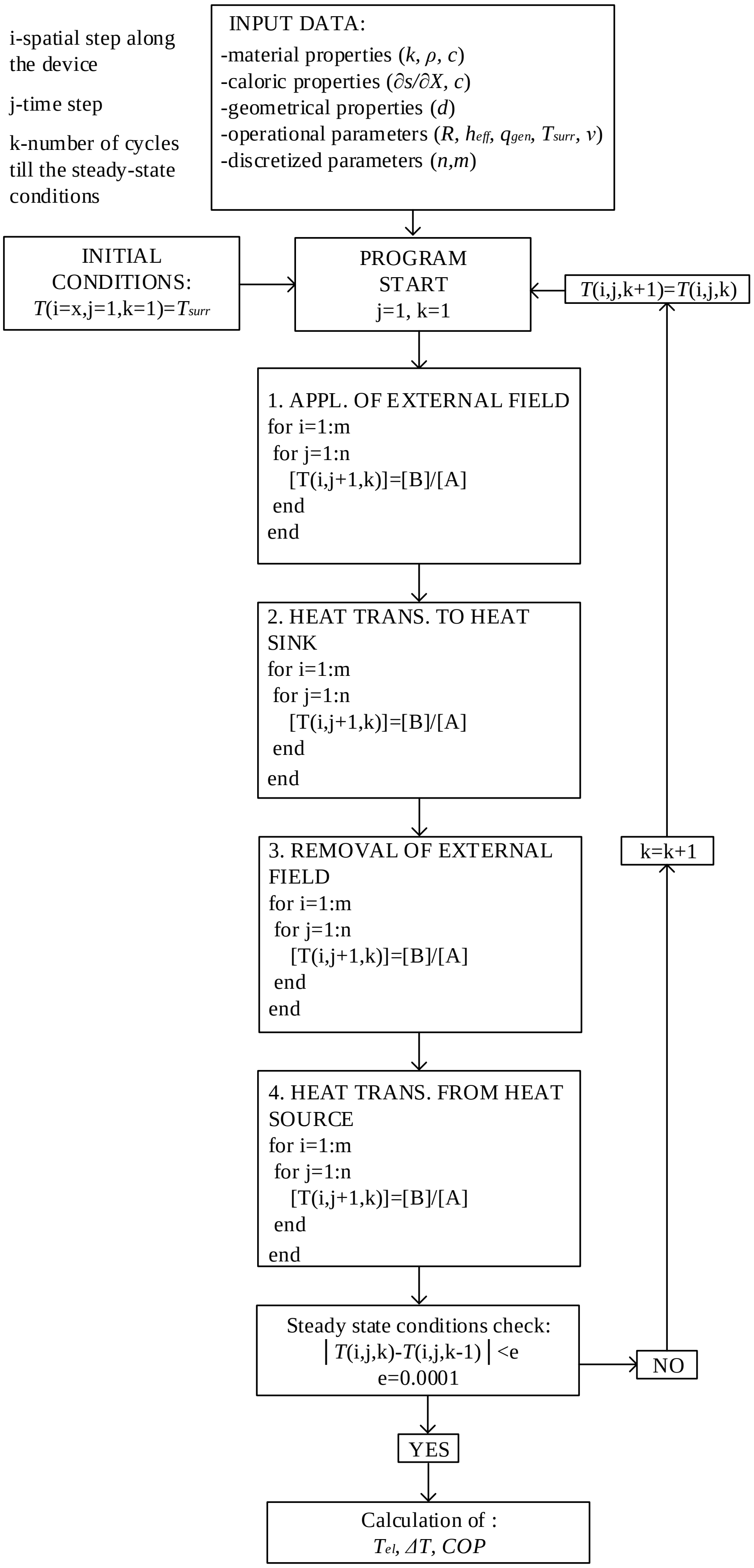

Figure 5: Flowchart of the developed numerical model.

## 3. Results and discussion

Figure 6a shows the temperatures of the Ni-Ti elastocaloric plate, the heat sink and the heat source (the electronic component) during the operation of the micro-cooling device from the initial condition (where the temperature of the entire device was at 290 K) until steady-state conditions were reached. The results shown in Figure 6 were obtained at an applied strain of 6 %, the contact thermal resistance was set to $2 \cdot 10^{-4}$ $m^2K/W$ and the effective heat-transfer coefficient to 300 $W/m^2K$. The thickness of the Ni-Ti plate was 1 mm and the operating frequency was 0.24 Hz. The heat flux of the electronic component was set to 0.3 $W/cm^2$. For a comparison, the temperature of the electronic component without active cooling (caloric plate) is also shown. For that purpose, the model was modified – it was assumed that an extended surface (fins) is mounted directly on the electronic component, while the other parameters remained the same. We can see that in both cases the steady-state conditions are approached after about 10 minutes. By applying a caloric micro-cooling device, the temperature of the electronic component is significantly lower (in this case by more than 10 K) compared to the case without the active cooling, but on the expanses on the additional work ($w_{in}$ – see Eq. (17)) consumed by a caloric device. Since the input work applied into the device ($w_{in}$) is converted to the heat, which needs to be transferred to the ambient (through the heat sink), this heats up the heat sink and, therefore, as seen from the Figure 6a, the hot side temperature ($T_{hot}$) of the caloric device is higher compared to the temperature of the electronic component if only passive cooling is applied ($T_{passive}$). Figure 6b further shows the temperatures of the elastocaloric plate and the heat sink/source during the initial cooling cycles. We can see that due to the larger thermal mass of the heat sink/source the temperature variation is much smaller compared to the elastocaloric plate, which impacts on the time required for the steady-state conditions to be reached.

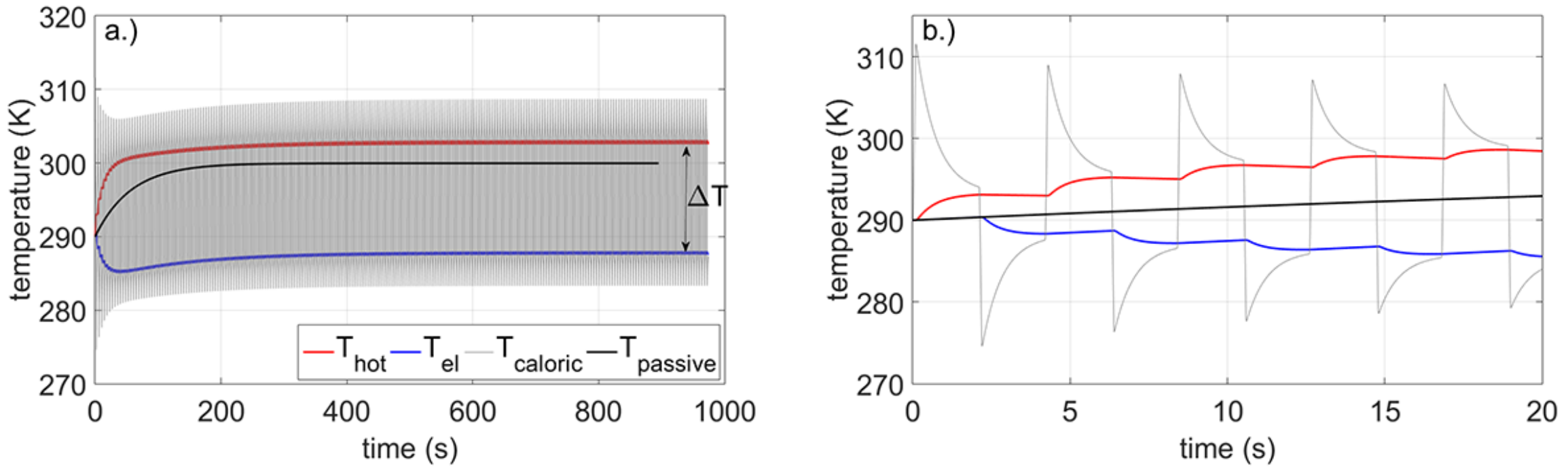

Figure 6: Temperature variations of the elastocaloric plate ($T_{caloric}$), the heat sink ($T_{hot}$) and the electronic component ($T_{el}$) during the operation of the micro-cooling device from the initial until the steady-state conditions (a); the temperatures of the elastocaloric plate and the heat sink and heat source during the initial five cooling cycles (b). For a comparison the temperature of the electronic component ($T_{passive}$) if only passive cooling (without the caloric plate) were to be applied is also shown.

### *3.1. Impact of heat transfer at the system's boundaries*

Prior to the investigation of the geometrical and operating conditions of the caloric micro-cooling device, the impacts of the effective heat-transfer coefficient of the heat sink and the contact thermal resistance between the caloric plate and the heat sink/source on the temperature of the electronic component and the COP were evaluated for different specific heat-flux densities (see Figure 7). All the other parameters remained unchanged (the caloric plate was Ni-Ti with an applied strain of 6%, its thickness was 1 mm and the operating frequency was 0.24 Hz). As seen from Figure 7, the temperature of the electronic component ($T_{el}$) is increasing with the decreased effective heat-transfer coefficient and also with the increased contact thermal resistance. If the effective heat-transfer coefficient in the heat sink is too low (below 200 W/m$^2$K for the heat fluxes applied here) the entire system would be overheated, since the generated heat would not be effectively transferred to the surroundings. Similar overheating of the micro-cooling device happens if the contact thermal resistance between the caloric plate and the heat sink/source is too high (above $5 \cdot 10^{-4}$ $m^2K/W$). On the other hand, as seen from Figures 7a and 7c the COP is almost independent of the effective heat-transfer coefficient and the contact thermal resistance. Even though the temperature of the whole system is increased at small values of the effective heat transfer coefficient and high values of the contact thermal resistance, this almost does not the COP values, since the caloric effect does not show significant temperature dependency (at the temperatures and the external fields applied in this study) as shown in Figure 4. As a comparison, Figure 7b also shows the temperature of the electronic component when no active cooling is applied (heat sink mounted directly on the electronic component without a caloric plate). It is evident that due to active cooling mechanism the caloric cooling provides significantly lower temperatures compared to forced convection only.

For further simulations, the effective heat-transfer coefficient was set to 300 $W/m^2K$, which is a realistic value for the air-cooled heat sink with forced convection and a fin effectiveness of 3.5 [3], [47]. The contact thermal resistance was set to $2 \cdot 10^{-4}$ $m^2K/W$, which normally occurs on smooth surfaces with a moderate contact pressure [47].

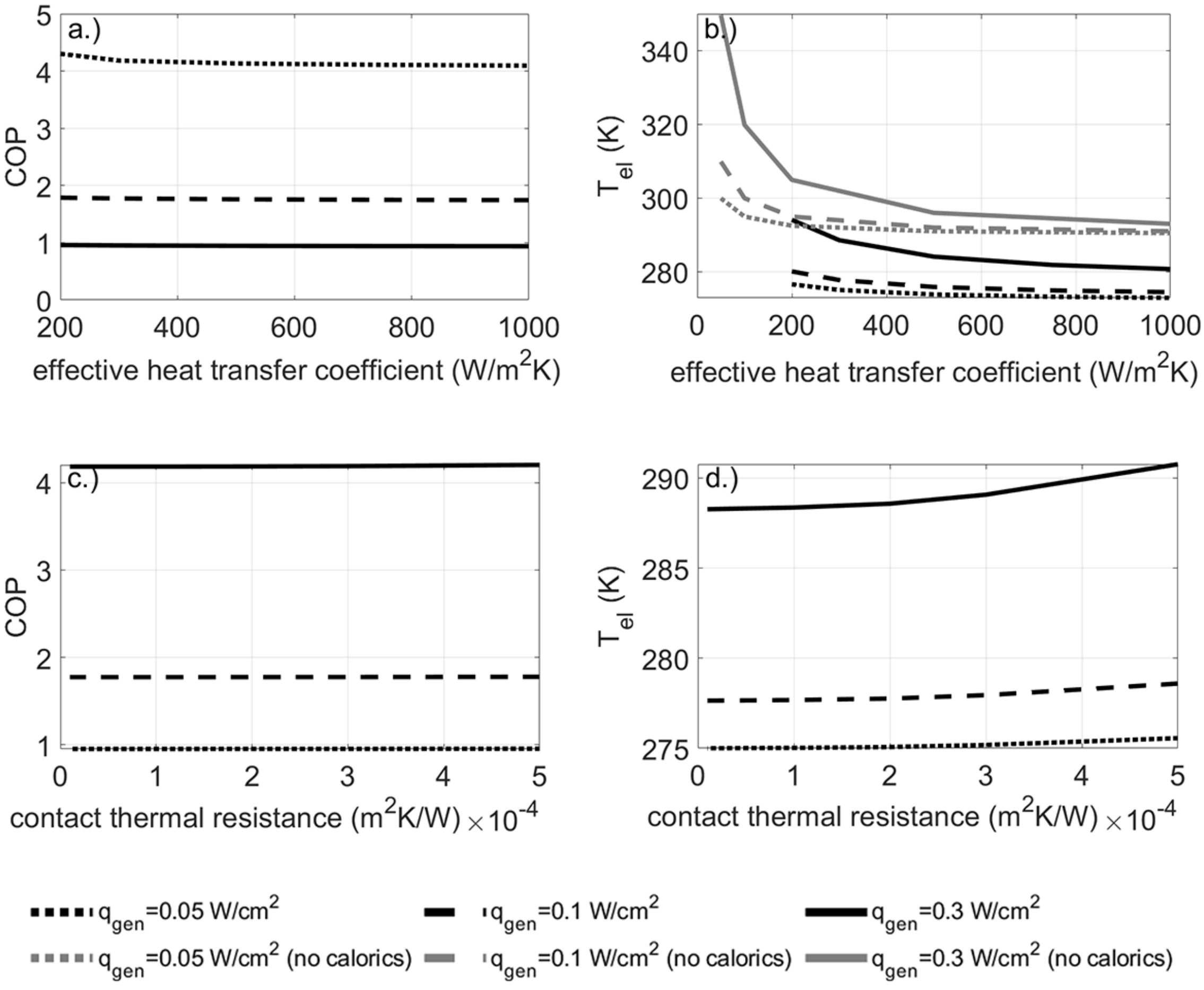

Figure 7: The electronic component's maximum temperature (a and c) and COP values (b and d) as a function of the effective heat-transfer coefficient (a and b) and the contact thermal resistance (c and d) for three different heat generations.

### *3.2.Impact of the caloric plate's thickness and operating conditions*

The impacts of the caloric plate's thickness and the operating conditions were evaluated for the Ni-Ti elastocaloric plate with an applied strain of 6%. Figure 8 shows the electronic component's maximum temperature, the temperature span between the heat sink and the heat source (calculated based on their average temperature), and the COP values as a function of the operating frequency for different thicknesses of the elastocaloric plate in the case of a heat-flux generation of 0.1 W/cm$^2$. We can see that from the temperature span point of view there is an optimal operating frequency for each plate thickness. The thicker the plate, the longer the heat-transfer time required for the heat to be transfer to the heat sink/source, which results in lower optimal operating frequency. Namely, there is an optimal heat-transfer time for each plate thickness. If the heat-transfer time is too short the latent heat of the caloric effect cannot be fully absorbed from the heat source, but if it is too long the temperature of the caloric plate reached the heat source temperature prior the end of the heat transfer period and, thus, the electronic component is not being actively cooled at this time period of the cooling cycle. However, the lowest temperature of the electronic component reached is about 278 K (with the largest temperature span being about 18 K), regardless of the plate thickness and the operating frequency. Namely, the largest temperature span is limited by the adiabatic temperature change of the caloric material and cannot be increased over this value; however, it is reached at a different operating frequency (for each plate's thickness). On the other hand, the COP values are decreasing with the operating frequency (for all the plate's thicknesses) and increasing with a decreased plate thickness (for all the operating frequencies). This is due to the fact that the input work is increasing with increased frequency and plate thickness (see Eq. (17)), while the specific heat-flux generation remained the same (0.1 W/cm$^2$). For the case shown in Figure 8, the COP values exceed 5 at an operating frequency below 0.5 Hz and a plate thickness of 0.1 mm. Nevertheless, the results indicate that the thickness of the caloric plate should be further decreased (below 0.1 mm) in order to additionally increase the COP values of the system. Similar trends of dependency were also previously shown for the operation of regenerative caloric devices [20].

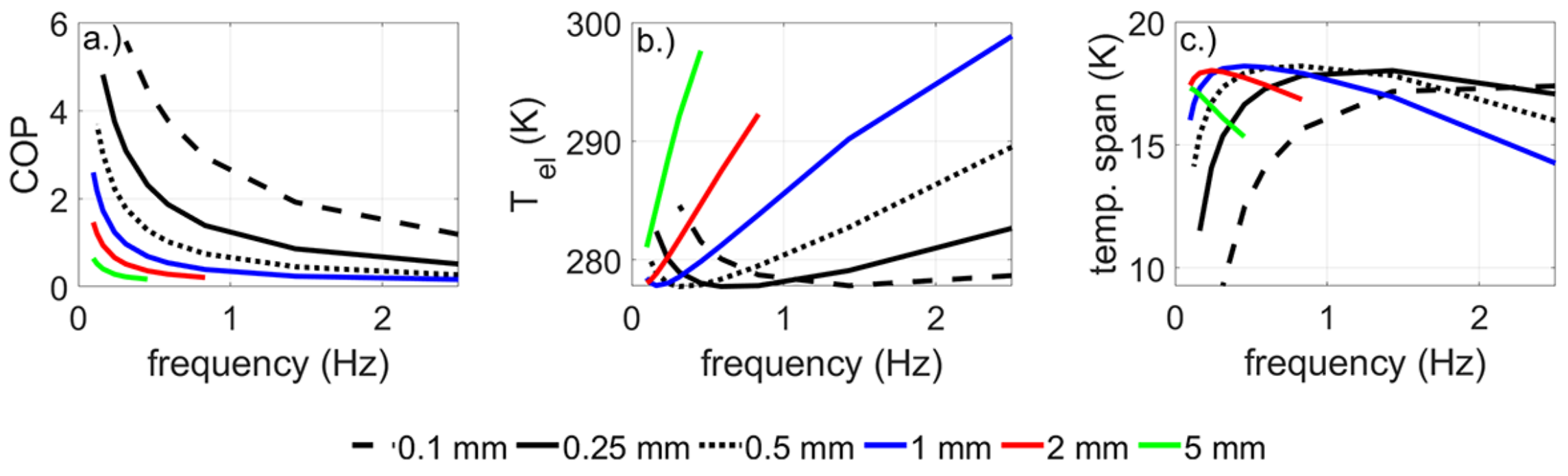

Figure 8: COP values (a), the maximum temperature of the electronic component (b) and the temperature span between the heat sink and heat source (c) as a function of operating frequency for different thicknesses of the elastocaloric plate in the case of a heat-flux generation of 0.1 W/cm$^2$.

### *3.3.Impact of the caloric material*

The same evaluation of the caloric plate's thickness and the operating conditions as shown above for the Ni-Ti elastocaloric plate with an applied strain of 6% was also performed for the other caloric materials considered in this study (see Figures S3-S13 in the Appendix A. Supplementary for the results). The main results are summarized in Figure 9, which shows the COP values and the temperature of the electronic component as a function of the specific cooling power (heat flux generation) for all four evaluated caloric materials under the selected (close-to-optimal) operating conditions from both the cooling power (heat-flux generation) and the COP point of view. We can see that there is a linear dependence of both the COP values and the temperature of the electronic component on the heat-flux generation. It is evident from Figure 9 that due to the larger caloric effect the Ni-Ti elastocaloric plate transfers higher heat-flux generation values compared to electrocaloric materials, but the COP values are generally smaller due to the larger hysteresis losses. For comparison, Figure 9b also shows the temperature of the electronic component with no active cooling (with heat sink mounted directly on the electronic component) as a function of the specific heat-flux generation. The application of a caloric cooling system is generally justified if it is capable of cooling the electronic component below room temperature or at least stabilizing its temperature at lower temperatures compared to passive cooling only. Figure 9b also shows a linear extrapolation of the electronic component's temperature as a function of heat-flux generation with only passive cooling (no calorics) and for the Ni-Ti elastocaloric plate strained by 6 %. We can see that the Ni-Ti strained by 6 % would be able to stabilize the electronic component at a lower temperature compared to only passive cooling for a heat-flux generation of up to 1 W/cm$^2$, while the electronic component would be cooled below room temperature up to a specific heat-flux generation of about 0.4 W/cm$^2$. If a smaller strain (2%) is applied to the Ni-Ti plate, the specific cooling power is lower compared to the strain of 6 % (see Figure 9b) due to smaller caloric effect (see Figure 4), but the COP values are higher (above 10), which is due to smaller hysteresis losses. On the other hand, the electrocaloric polymer can work with heat-flux generation values up to about 0.05 W/cm$^2$ and COP values up to 9, while the electrocaloric ceramic can work with heat flux generation values up to 0.03 W/cm$^2$, but with the highest COP values (above 30), owing to the smallest hysteresis losses.

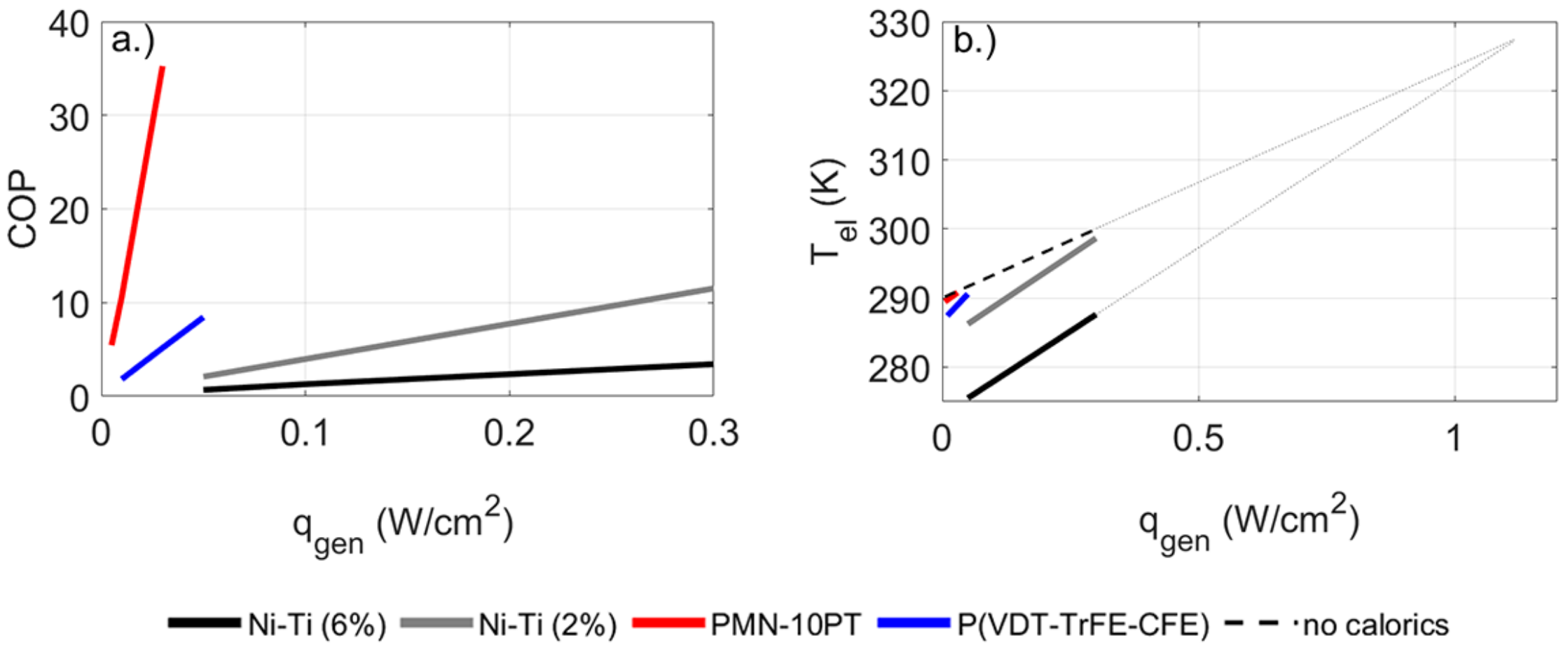

Figure 9: Maximum temperature of electronic component (a) and COP values (b) for different caloric materials (for a caloric plate thickness of 0.5 mm and an operating frequency of 0.45 Hz) as a function of heat-flux generation.

Since our own experimental data are not yet available, the data form the existing literature [40], [45], [60] were used for the basic validation of the model. Due to somewhat different working conditions the direct comparison was not possible, but quantitatively our modelling results fit relatively well with compared data from the literature. For example, the electrocaloric device developed by Ma et al. [40] that employed 0.06 mm thick P(VDT-TrFE-CFE) polymer film as the caloric material at the electric field change of 667 kV/cm and operating frequency of 0.8 Hz generated specific cooling power of around 0.03 W/cm$^2$ with the COP of 13 at the temperature span of 1.4 K. Our modelling results show that 0.1 mm thick P(VDT-TrFE-CFE) polymer film at the electric filed change of 750 kV/cm and operating frequency of 0.83 Hz generates specific cooling power (heat-flux) of 0.03 W/cm$^2$ with the COP of about 12.5 at the temperature span of about 2.3 K (see Figure S9 in in the Appendix A. Supplementary data). Even though the operating conditions are not exactly the same, the modelling and experimental results show good agreement, while the observed deviations can be attributed to the differences in contact heat transfer resistance and effective heat transfer coefficient of the heat sink between the model and the device.

## 4. Conclusions

A new, transient numerical model for simulating and optimizing the operation of a caloric micro-cooling device was developed and implemented in Matlab software. The micro-cooling characteristics of three different caloric materials (a Ni-Ti alloy strained by 6 % and 2 %, a PMN-10PT relaxor ceramic and a P(VDT-TrFE-CFE) polymer) were evaluated and investigated/optimized. Initially, we demonstrated the importance of the contact thermal resistance and in particular the convective heat-transfer coefficient of the heat sink on the performance of the caloric micro-cooling device. If the convective heat-transfer coefficient is too low, the whole system would overheat, but if the convective heat-transfer coefficient is increased, the caloric plate can help to cool down the electronic component to lower temperatures (compared to the case when only the heat sink is applied) and even cool it below the room temperature. We have further shown the importance of the operating conditions (frequency and caloric plate thickness) on the performance of the caloric micro-cooling device. It works more efficiently (with higher COP values) at lower frequencies and smaller thicknesses of the caloric plate. On the other hand, from the temperature-span point of view there is an optimal operating frequency for each plate thickness. Nevertheless, it was shown that caloric micro-cooling systems in combination with air-cooled heat sinks are able to under cool electronic components below room temperature at relatively small heat fluxes (up to 0.05 $W/cm^2$ for electrocaloric elements and up to 0.35 $W/cm^2$ for elastocaloric elements), but with relatively high COP values (up to around 10 or even more if a PMN-10PT relaxor-ceramic electrocaloric element is used). It is further shown that applying a caloric micro-cooling device allows lowering of the electronic component's temperature (compared to the case where only air-cooled heat sinks with no calorics are used) up to heat flux generations of 1 $W/cm^2$. This value can be significantly increased if instead of air-cooled heat sinks, more powerful heat sinks, such as water-cooled or even heat pipes, were to be applied. Although not shown here in details, the modelling results revealed that if, for example, the effective heat-transfer coefficient would be 1000 $W/m^2K$ (normal values for water-cooled heat sinks), the Ni-Ti plate strained by 6 % would be able to cool down the electronic component below room temperature up to heat-flux generation values of about 1 $W/cm^2$. The caloric micro-cooling device can therefore additionally reduce the temperature (even below room temperature) of an electronic component in an efficient way as long as they are applied in combination with powerful heat sinks. This could be a significant advantage in several hot-spot cooling and similar applications that require efficient cooling with heat-flux densities of up to around 1 $W/cm^2$ in the future. However, in order for the caloric micro-cooling technology to be applied in the applications with higher values of heat-flux densities (well above 1 W/cm$^2$) significant improvements should be made in the development of new caloric materials and well as in the concepts of caloric micro-cooling devices.

### CRediT authorship contribution statement

**Jan Kalizan**: Software, Validation. **Jaka Tušek**: Supervision, Conceptualization, Methodology, Visualization, Writing - original draft, Writing - review & editing.

### Declaration of Competing Interest

The authors declare that they have no known competing financial interests or personal relationships that could have appeared to influence the work reported in this paper.

### Acknowledgements

The financial support of Slovenian Research Agency (project nr. J2-1738) is highly appreciated. Jaka Tušek would like to acknowledge financial support of European Research Council (ERC) under Horizon 2020 research and innovation program (ERC Starting Grant No. 803669).

### Appendix A. Supplementary data